\documentclass[11pt]{article}
\usepackage{moriond,epsfig,subfigure}

\usepackage{fancybox}%         e.g. for calling out requirements, etc.             
\usepackage{multicol}%         multiple columns (e.g. glossary)

\usepackage{fancyhdr}%        modern version of fancyheadings
\usepackage{subfigure}%
\usepackage{xspace}% 
\usepackage{wrapfig}% 

\usepackage{amsmath,amssymb}% for advanced math typesetting
\usepackage{topcapt}% for captions at top, i.e., tables

\def\be{\begin{equation}}
\def\ee{\end{equation}}
\def\bea{\begin{eqnarray}}
\def\eea{\end{eqnarray}}

\newcommand{\cm}{\ensuremath{\,\text{cm}}\xspace}

\newcommand{\GeV}{\ensuremath{\,\text{Ge\hspace{-.08em}V}}\xspace}
\newcommand{\TeV}{\ensuremath{\,\text{Te\hspace{-.08em}V}}\xspace}

\newcommand{\GeVc}{\ensuremath{{\,\text{Ge\hspace{-.08em}V\hspace{-0.16em}/\hspace{-0.08em}}c}}\xspace}

\newcommand{\GeVcc}{\ensuremath{{\,\text{Ge\hspace{-.08em}V\hspace{-0.16em}/\hspace{-0.08em}}c^\text{2}}}\xspace}

\newcommand{\pbinv} {\mbox{\ensuremath{\,\text{pb}^\text{$-$1}}}\xspace}

\newcommand{\ttbar}{\ensuremath{{t\overline{t}}}\xspace}
\newcommand{\ttbarjets}{\ensuremath{{t\overline{t}}+jets}\xspace}
\newcommand{\Znunujets}{\ensuremath{{\rm Z}\rightarrow \nu \nu}+jets\xspace}
\newcommand{\Zmumu}{\ensuremath{{\rm Z}\rightarrow \mu \mu}\xspace}
\newcommand{\Zee}{\ensuremath{{\rm Z}\rightarrow e e}\xspace}
\newcommand{\Wjets}{W+jets\xspace}
\newcommand{\Zjets}{Z+jets\xspace}

\newcommand{\pt}{\ensuremath{p_{\mathrm{T}}}\xspace}

\newcommand{\HT}{\ensuremath{H_{\mathrm{T}}}\xspace}
\newcommand{\MHT}{\ensuremath{H_{\mathrm{T}}^{\text{miss}}}\xspace}

\newcommand{\MET}{\ensuremath{E_{\mathrm{T}}^{\text{miss}}}\xspace}

\newcommand{\rpv}{\ensuremath{\rlap{\kern.2em/}R}\xspace}

\begin{document}
\vspace*{4cm}
\title{Searches for Supersymmetry at CMS using the 2010 Data}

\author{C. Bernet, on behalf of the CMS collaboration}

\address{CERN, Geneva.}

\maketitle\abstracts{
Searches for supersymmetry were conducted using the 35 ${\rm pb}^{-1}$ of data collected by the CMS experiment at the LHC in 2010, at a centre-of-mass energy of 7~\TeV. A wide variety of final states featuring jets and missing transverse energy, possibly with leptons, were investigated. 
The data, consistent with the standard-model hypothesis, allow us to set limits on the existence of new physics, 
extending those previously obtained at the Tevatron and LEP. 
}

\section{Introduction}

The standard model (SM) of particle physics has been enormously successful in describing all phenomena at the highest attainable energies thus far. Yet, it is widely believed to be only an effective description of a more complete theory which is valid at the highest energy scales. Of particular theoretical interest is supersymmetry (SUSY)~\cite{ref:SUSY0,ref:SUSY1,ref:SUSY2,ref:SUSY3,ref:SUSY4} 
which solves the hierarchy problem~\cite{ref:hierarchy1,ref:hierarchy2} of the SM by compensating for each of the fermionic and bosonic degrees of freedom in the SM with a supersymmetric bosonic and fermionic degree of freedom, respectively. The resulting superfields have the same quantum numbers as their SM counterparts, except for spin. 
%a fermionic (bosonic) degree of freedom for each bosonic (fermionic) degree of freedom in the SM. 
%large number of supersymmetric particles with the same quantum numbers as the SM particles, but differing by half a unit of spin.  %wherein an artificial cancellation of very large terms is required for %the Higgs boson mass to be at the electroweak scale.  %Moreover, in SUSY models with R-parity conservation, the lightest supersymmetric %particle (LSP) is stable and is an excellent candidate for dark %matter~\cite{ref:SUSY-LSP-DM}. If R-parity conservation is assumed, supersymmetric particles are produced in pairs, and decay to the lightest supersymmetric particle (LSP), leading to the celebrated missing energy signature of supersymmetry. 
Since no SUSY particle has been observed so far, they must have higher masses than their SM partners, implying that SUSY is a broken symmetry. 
%, and therefore SUSY must be a broken symmetry.  %There are numerous models for SUSY breaking which maintain all advantages of a supersymmetric theory.  A leading and practical model of SUSY-breaking is the Constrained Minimal Supersymmetric extension of the Standard Model (CMSSM)~\cite{ref:CMSSM} described by five parameters: the universal scalar and gaugino mass parameters ($m_0$ and $m_{1/2}$, respectively); the universal trilinear coupling ($A_0$); %all of which are assumed to be equal at some Grand-Unification scale, and two low-energy parameters, the ratio of the two vacuum expectation values of the two Higgs doublets, $\tan\beta$, and the sign of the Higgs mixing parameter, $\sign(\mu)$. Experiments at the energy frontier, i.e. at the Fermilab Tevatron Collider~\cite{CDFLimits,D0Limits,Abazov200934}, and previously at the CERN SPS~\cite{UA1Limits,UA2Limits} and LEP Colliders~\cite{LEPLimits} have performed extensive searches for signs of SUSY. The experiments have observed no such signs and have set lower limits on the masses of SUSY particles. 

At the Large Hadron Collider (LHC) at CERN, 
supersymmetric particles, if they exist, are predicted to be produced dominantly via QCD, through the fusion of 
two gluons into a pair of gluinos, a pair of squarks, or a gluino and a squark. 
The production cross-section for massive squarks or gluinos falls as a power law with the squark or gluino mass,  following the available energy $\sqrt{\hat s}$ in the partonic centre-of-mass frame. 
The LHC, with a proton-proton centre-of-mass energy $\sqrt{s}$ of 7~\TeV, is a copious source of high-energy partons which allows to probe squark and gluino masses beyond the limits previously set at LEP and at the Tevatron. 
Squarks and gluinos initiate a decay cascade in which quarks are produced, until the lightest supersymmetric particle (LSP) is created. 
The dynamics of the cascade depends on the SUSY model under consideration, and in particular on the masses of the supersymmetric particles. 
If R-parity is conserved, the LSP is unable to decay into SM particles and is therefore stable. 
If, in addition, the LSP is a neutralino, it is weakly interacting and thus escapes detection, hence missing transverse energy (\MET) in the final state. 
Typical hadronic decay modes for gluinos ($\tilde g$) and squarks ($\tilde q$) are $\tilde q \rightarrow q \chi_1^0$ and  $\tilde g \rightarrow q q \chi_1^0$.
In these examples, the squark and the gluino directly decay to the lightest neutralino $\chi_1^0$, the gluino doing so via an off-shell squark. As a result, squark pair production usually gives rise to more than two jets, and gluino pair production to more than four jets. The transverse momenta of the jets are driven by the difference in mass between the squark or gluino and the neutralino. 
%The amount of missing transverse energy arising from the presence of the neutralino is not directly related to the neutralino mass. 
%
Leptons can appear in the final state, for example if heavy neutralinos ($\tilde  \chi_2^0 \rightarrow l^\pm \tilde l^\mp \rightarrow l^\pm l^\mp \chi_1^0 $) or charginos ($\tilde  \chi_1^\pm \rightarrow \chi_1^0 W^\pm$) are created in the decays cascades of the squark or gluino.

The CMS detector~\cite{ref:CMS} is used to investigate many final states that could arise from the strong production of squarks and gluinos. An effort is made to make these final states independent, so that all analyses can ultimately be easily combined.  
Because the presence of leptons is not guaranteed, investigating hadronic final states with jets and high missing transverse energy is the most efficient way to look for SUSY.  
Dealing with the huge QCD background is however a challenge. 
In CMS, three complementary approaches are followed. The $\alpha_T$ analysis, presented elsewhere~\cite{ref:CMSalphaT} makes use of the $\alpha_T$ variable to completely remove the QCD background from the search region, leaving solely electroweak backgrounds, namely \ttbarjets, \Wjets and \Znunujets. 
The jets + \MHT analysis, summarized in Section~\ref{sec:RA2}, consists of looking for an excess of multi-jet events at high \MHT, an approximation of the \MET computed as the opposite of the vector sum of the jet transverse momenta.  
This approach is the most efficient of the three, but requires the QCD background to be accurately controlled. 
The so-called razor analysis, presented in Section~\ref{sec:Razor} relies on novel variables to reduce the QCD background to a negligible level in the search region, and to predict the background contribution. 
The final search sample of this analysis has about 30\% of events in common with the jets+\MHT analysis. While the razor analysis is less efficient than the jets+\MHT analysis, it is less sensitive to the effects of initial state radiation. Requiring leptons in the final state in addition to jets and missing transverse energy strongly reduce the standard-model background. 
With one isolated lepton, the QCD and \Znunujets backgrounds get suppressed.  
With two opposite-sign leptons~\cite{ref:CMSOppositeSign}, the \Wjets background becomes negligible, and several handles can be used for an accurate estimation of the remaining \ttbar background from the data. Asking for two same-sign leptons, or for three or more leptons, dramatically suppresses the standard-model background, for a very clean search of physics beyond the standard model, like the production of squarks and gluinos which can naturally lead to such final states.

In these proceedings, the emphasis is put on the most recent fully hadronic analyses, and several leptonic analyses are briefly summarized. Other important search fields are also being covered by CMS but could not be presented here. For example, in the context of the general gauge-mediated SUSY breaking with the lightest neutralino as the next-to-lightest supersymmetric particle and the gravitino as the lightest, a natural signature for squark or gluino production is the presence of two photons and \MET in the final state~\cite{ref:CMSDiPhoton}. 

%A detailed description of the Compact Muon Solenoid (CMS) detector can be found elsewhere~\cite{ref:CMS}. 
%A characteristic feature of the detector is its superconducting solenoid magnet, of 6~m internal diameter, providing a field of 3.8~T. The silicon pixel and strip tracker, the crystal electromagnetic calorimeter (ECAL) and the brass/scintillator hadron calorimeter (HCAL) are located within the solenoid. Muons are detected in gas-ionization detectors embedded in the steel return yoke outside the magnet. 
%The ECAL has an energy resolution of better than 0.5\,\% above 100~GeV. The HCAL, when combined with the ECAL, measures jets with a resolution $\Delta E/E \approx 100\,\% / \sqrt{E} \oplus 5\,\%$. 
%The particle flow algorithm~\cite{PFT-09-001,PFT-10-002} identifies and reconstructs all particles produced in the collision, namely charged hadrons, photons, neutral hadrons, muons, and electrons. The resulting list of particles is then used to e.g. reconstruct particle jets, compute the \MET, and quantify the lepton isolation.

%CMS uses a right-handed coordinate system, with the origin located at the nominal collision point, the $x$-axis pointing towards the centre of the LHC, the $y$-axis pointing up (perpendicular to the LHC plane), and the $z$-axis along the (anti-clockwise) beam direction. The azimuthal angle $\phi$ is measured with respect to the $x$ axis in the $xy$ plane and the polar angle $\theta$ is defined with respect to the $z$ axis. The pseudorapidity is $\eta = -\ln (\tan(\theta / 2) )$.

\section{Fully hadronic searches}

\subsection{Jets + \MHT analysis}
\label{sec:RA2}

The data used in this analysis~\cite{ref:CMSRA2} are collected using triggers requiring a minimal jet activity $\HT^{\rm trig}$, measured as the scalar sum of the transverse momentum of the calorimeter jets reconstructed at trigger level. 
The rapid increase in instantaneous luminosity during the 2010 data taking resulted in the threshold on $\HT^{\rm trig}$ being raised from $100$ to $140$ and finally $150 \GeV$. 
%The choice of an \HT trigger aligns with the requirements above for the event selection. It provides inclusiveness with respect to a hadronic signal, a high acceptance for low-mass hadronic new-physics signatures, and it enables the simultaneous collection of several control samples used to estimate backgrounds to the search.
The particle flow algorithm~\cite{PFT-09-001,PFT-10-002} identifies and reconstructs all particles produced in the collision, namely charged hadrons, photons, neutral hadrons, muons, and electrons. The resulting list of particles is then used to reconstruct particle jets, compute the \MET, and quantify the lepton isolation.

The event selection starts from a loose validation region. On top of this baseline selection, tighter selection criteria are applied to define the search regions. The baseline selection requirements after trigger boil down to selecting events with (i) at least three jets with $\pt> 50\GeVc$ and $| \eta | < 2.5$; (ii) $\HT > 300 \GeV$, where \HT is  defined as the scalar sum of the transverse momenta of all the jets having $\pt>50 \GeVc$ and $|\eta|<2.5$;  (iii) $\MHT> 150\GeV$, where \MHT is defined as the magnitude of the vectorial sum of the \pt of the jets having, in this case, $\pt>30 \GeVc$ and $| \eta | < 5$; (iii) no isolated electron nor muon with $\pt > 10\,\GeVc$. Additionally, the \MHT vector is required not to be aligned with one of the three leading jets, to reject QCD multi-jet events in which a single mis-measured jet yields high \MHT. 

Two search regions were defined in this inclusive jets-plus-missing-momentum search. The first search selection, defining the ``high-\MHT search region'', tightens the baseline cuts with an $\MHT > 250\GeV$ requirement, to search for a generic invisible particle in a low background environment. The second selection adds a $\HT > 500\GeV$ cut to the baseline selection, yielding the ``high-\HT search region'', sensitive to cascade decays of high-mass new-physics particles where more energy is transferred to visible particles and less to the dark-matter candidate.
The main background contributions in the two search regions are estimated using data-driven techniques. 
Due to its huge cross-section, QCD multi-jet production can give rise to high \MHT because of the finite jet energy resolution, or of rare but dramatic mis-measurements of the jet energy induced by various instrumental effects.
The most important instrumental effects were identified in the simulation to be related to missing channels in the ECAL, and to jet punch-through giving rise to multi-TeV fake muons in the particle jets. The simulation was used to design dedicated event filters to remove such events. 

The QCD background was estimated using the so-called ``rebalance+smear'' technique. An inclusive multi-jet sample of events is selected. The energy of each jet is first rescaled to obtain a null \MHT using a maximum-likelihood fit taking into account the jet energy resolution in the process. This rescaling produces a seed event from which all sources of \MHT, possibly genuine, have been removed.  
The jets are then smeared by a simulated jet energy response distribution. The simulated distribution is corrected for differences between the data and the simulation by factors obtained from di-jet asymmetry measurements. 
The other standard-model background events contributing to the search regions feature at least one neutrino in the final state, hence true \MHT. 
\Wjets events, where the W possibly comes from a top and decays to a lepton and a neutrino, end up in the search region in case the lepton from the W decay is not identified in the analysis, either because it is a $\tau$ decaying hadronically, or an electron or muon that is lost (not identified by the lepton veto). 
The contribution of this source of background is estimated by selecting from the data a control sample of events with an isolated muon and jets. 
To predict the number of events with a lost lepton in the search region, the number of events in this control sample is corrected for lepton reconstruction and identification efficiency by factors measured using $Z$ events in the data, and by acceptance factors from the simulation.  
To estimate the number of events in which a tau decays hadronically, the muon in the control sample is replaced by a jet representing the hadronically decaying tau, which is taken into account when applying the search selections. 
The uncertainty on the \Wjets background estimation (including \ttbar) is dominated by the statistical error on the number of events in the control sample. 
The last source of background, especially important because it dominates at high \MHT, is \Znunujets.
As no \Zee nor \Zmumu events are observed in the search regions, these processes cannot be used to predict the \Znunujets background contribution. This contribution is instead estimated using a control sample of isolated $\gamma+$jets events, in which the photon is ignored when applying the search selections. This strategy exploits the fact that at high boson \pt, the Z and $\gamma$ behave in a similar way, apart from electroweak coupling differences, and small residual mass effects. 
The number of events in the control sample is corrected by a $Z/\gamma$ cross-section correction factor obtained from the simulation. Several other effects, such as the contamination of the control sample by multi-jet QCD events, or the photon reconstruction and identification efficiencies, are taken into account. The error on this background prediction comes from the statistical error on the number of events in the control sample, and from systematic errors mostly related to the available number of events in the simulated samples and to the estimation of the contamination of the control sample by multi-jet events. Table~\ref{tab:RA2} summarizes the results of the analysis, and shows that no excess of events is found in the data. The limit set on the number of signal events is interpreted in the context of various SUSY models in Section~\ref{sec:ModelInterpretation}.

\begin{table}[htb]
\begin{center}
\caption{Predicted and observed event yields for the baseline selection, and for the high-\MHT and high-\HT search selections. 
% The total background estimate from data, used in the limit calculations, uses the QCD R+S, the \Znunujets from photons and the W/ \ttbar lost-lepton and hadronic-tau estimates. The background combination is performed as explained in the text. 
The last line reports the 95\% CL limit on the number of signal events given the observed number of events, and the total predicted background. 
}
\label{tab:RA2}
{
\begin{tabular}{|l|rr|rr|rr|}
\hline
  Background & \multicolumn{2}{c|}{Baseline} & \multicolumn{2}{c|}{High-\MHT} & \multicolumn{2}{c|}{High-\HT}  \\
         & \multicolumn{2}{c|}{selection} & \multicolumn{2}{c|}{selection} & \multicolumn{2}{c|}{selection} \\\hline
  $\Znunujets$ ($\gamma$+jets method)            & 26.3  & $\pm 4.8$         & 7.1  & $\pm 2.2$  & 8.4  & $\pm 2.3$ \\
  $W/\ttbar\to e,\mu$+X                    & 33.0  & $\pm 8.1 $        & 4.8  & $\pm 1.9 $ & 10.9 & $\pm 3.4$ \\
  $W/\ttbar\to \tau_{\mbox{\tiny hadr}}$+X  & 22.3  & $\pm 4.6 $        & 6.7  & $\pm 2.1 $ & 8.5  & $\pm 2.5$ \\
  QCD                          & 29.7  & $\pm 15.2$        & 0.16 & $\pm 0.10$ & 16.0 & $\pm 7.9$ \\
 \hline
  Total background estimated from data               & 111.3 & $\pm 18.5$        & 18.8 & $\pm 3.5$  & 43.8 & $\pm 9.2$ \\
  Observed in $36 \pbinv$ of data            & 111\hspace{3mm}   &     & 15\hspace{3mm} &  & 40\hspace{3mm} & \\
  \hline
  95\% C.L. limit on signal events                    & 40.4  &     & 9.6 & & 19.6 & \\
\hline
\end{tabular}
}
\end{center}
\end{table}

% The set of \HT triggers used for the 2010 data is found to be fully efficient for \HT above $300 \GeV$.
%, as shown in Fig.~\ref{fig:trigger_HTeff}. The use of particle-flow jets offline induces a slow efficiency turn-on.

% The trigger requires a minimal jet activity, measured as the scalar sum of the transverse energy of the calorimetric jets reconstructed at trigger level. 

\vspace{-0.7cm}
\subsection{Razor analysis}
\label{sec:Razor}

This analysis relies on the novel ``razor'' variables~\cite{ref:RazorRogan} to define search regions and predict the background contribution in a data-driven way. For the pair-production of two heavy particles of mass $M_{\tilde q}$ decaying into a visible part and an invisible part of mass $M_\chi$, the variable $M_R$ provides an approximation of the quantity
$M_\Delta \equiv  (M_{\tilde{q}}^{2}-M_{\tilde\chi}^{2})/M_{\tilde{q}}$.
The search consists of looking for a signal peak in the $M_R$ distribution, on top of a steeply falling standard-model background distribution. Cutting on the dimensionless $R$ variable strongly reduces the standard-model background, and to give its $M_R$ distribution an easy-to-control exponential shape. 
%Figure~\ref{fig:MRvR} shows the distribution of $R$ as a function of $M_R$, for the QCD multi-jet background and LM1, a SUSY benchmark model used in CMS, and described in Section~\ref{sec:ModelInterpretation}.
%
%\begin{figure}[ht!]
%\centering
%\includegraphics[width=0.49\textwidth]{Pictures/Razor/R-MR-QCD.pdf}
%\includegraphics[width=0.49\textwidth]{Pictures/Razor/R-MR-LM1.pdf}
%\caption{The razor plane: $M_R$ versus $R$ yields for 10 pb$^{-1}$
%  Monte Carlo simulated samples: QCD multi-jets (a) and a CMS SUSY benchmark model (LM1) (b) with $M_\Delta = 597$\,\GeV.}
%\label{fig:MRvR}
%\end{figure}
%

The razor analysis~\cite{ref:CMSRazor} defines a set of physics objects, namely jets, isolated electrons, and isolated muons. All of these objects are used in the computation of $R$ and $M_R$, which proceeds in the following way. The objects are first grouped into two ``mega-jets'' using an hemisphere algorithm. Each mega-jet ideally corresponds to the visible part of the decay products of one of the pair-produced heavy particles. The $R$ and $M_R$ variables are then computed using the 4-momenta of the two mega-jets, and the \MET vector. Depending on the presence of an isolated electron or muon in the final states, the events are classified in three independent ``boxes'', the electron box, the muon box, and the hadronic box. The high $R$, high $M_R$ region of each box constitutes an independent search region. The razor analysis is thus both a fully hadronic and a single lepton analysis. In these proceedings however, the focus is put on the more efficient fully hadronic sector, for which the low-$M_R$ region of the leptonic boxes is used as a control sample

\begin{wrapfigure}[16]{r}{3.5in}
\begin{center}
\vspace{-0.7cm}
\includegraphics[width=0.49\textwidth]{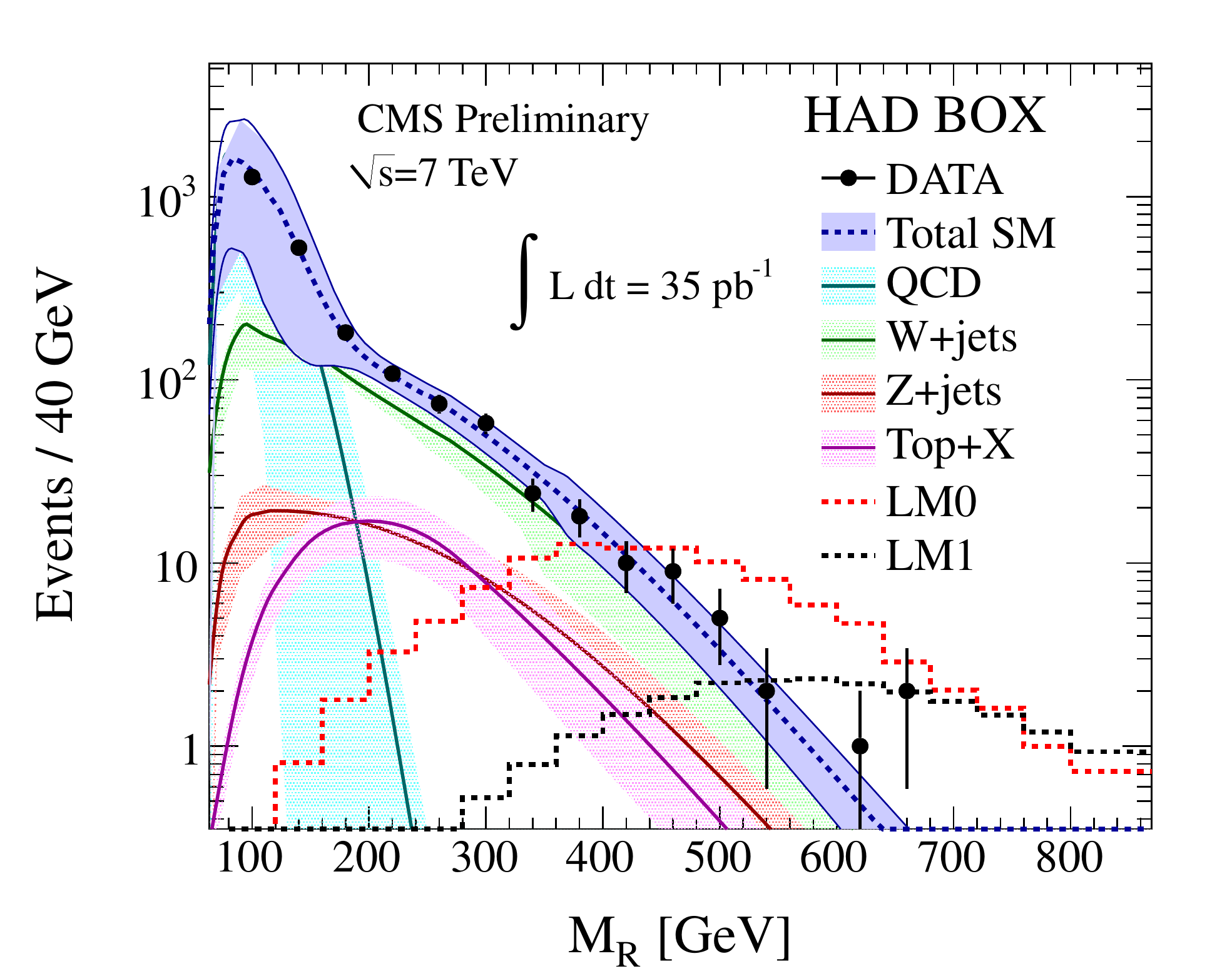} 
\caption{Distribution of $M_R$ in the data, and background prediction for the ``razor'' analysis, with $R > 0.5$. The search region is defined by the additional requirement $M_R > 500\,\GeVcc$. }
\label{fig:HADBOX}
\end{center}
\end{wrapfigure}
The event sample was collected using triggers based on the presence of a single electron, a single muon, and on $H_T^{\rm trig}$. The jets are required to have $\pt > 30$\,\GeVc and $|\eta|<3$, electrons to have $\pt > 20$\,\GeVc and $|\eta|<2.5$, and muons to have $\pt > 20$\,\GeVc and $|\eta|<2.1$. The difference in azimuth between the two mega-jets is required to be smaller than 2.8\,rad, to reject di-jet QCD events. 
The $M_R$ distribution for the data in the hadronic box, together with the full background prediction, is shown in Fig.~\ref{fig:HADBOX}, for $R>0.5$. 

The background prediction is based on the observation that above a given value of $M_R$, all background distributions drop following an exponential function. At low $M_R$, the background shape is mostly driven by the efficiency of the \HT trigger, and by the mass scales of the standard-model processes. For instance, $M_R$ peaks around the mass of the top quark for \ttbar events. 

For the \ttbar, \Znunujets, and \Wjets backgrounds, the parameters of the exponential function driving the evolution of the $M_R$ distribution at high $M_R$ are taken from the simulation. In the simulation, and also in Fig.~\ref{fig:HADBOX}, these parameters appear to be roughly equal, indicating a similar behaviour of these background processes in terms of $R$ and $M_R$. These parameters are then corrected by factors compatible with one, extracted from a comparison between data and simulation for \Wjets events in the muon box. The relative normalization of these three sources of background is set according to the inclusive W, Z, and \ttbar cross-sections measured by CMS~\cite{ref:CMSWZ,ref:CMSttbar}. The normalization of the overall background distribution to the data is performed by measuring lepton boxes event yields, which are then corrected for lepton reconstruction and identification efficiency. A fit is finally performed in the $80 < M_R < 400$\,\GeVcc region, to obtain the parameters of the \HT trigger turn-on shape and the overall normalization of the QCD background. The shape of the QCD background was obtained using a low-bias, prescaled trigger. The background is predicted by extrapolating the resulting background distribution to the search region, defined as $M_R>500\,\GeVcc$. In this region, 7 events are observed in the data, and $5.5\pm1.4$ are expected from the background. As no excess is observed, a model-independent upper limit is set on the number of signal events, $s<8.4$. This limit is interpreted in the context of various SUSY models in Section~\ref{sec:ModelInterpretation}.

\subsection{Model dependent interpretation}
\label{sec:ModelInterpretation}

The results of the fully hadronic (Sections~\ref{sec:RA2} and \ref{sec:Razor}) analyses were interpreted in the context of the constrained MSSM (cMSSM), a truncation of the full parameter space of the MSSM motivated by the minimal supergravity framework for spontaneous soft breaking of supersymmetry. In the cMSSM, the soft breaking parameters are reduced to five: three mass parameters, $m_0$, $m_{1/2}$ and $A_0$ being respectively the universal scalar mass, the universal gaugino mass, and the universal trilinear scalar coupling, as well as ${\rm tan} \beta$, the ratio of the up-type and down-type Higgs vacuum expectation values, and the sign of the supersymmetric Higgs mass parameter $\mu$. Scanning over this parameter space yields models which, while not entirely representative of the complete MSSM, vary widely in their supersymmetric mass spectra and thus in the dominant production channels and decay chains.

\begin{figure}[htb]
\centering
\hspace{-1.3cm}
\subfigure[]{\includegraphics[angle=0,width=0.47\textwidth]{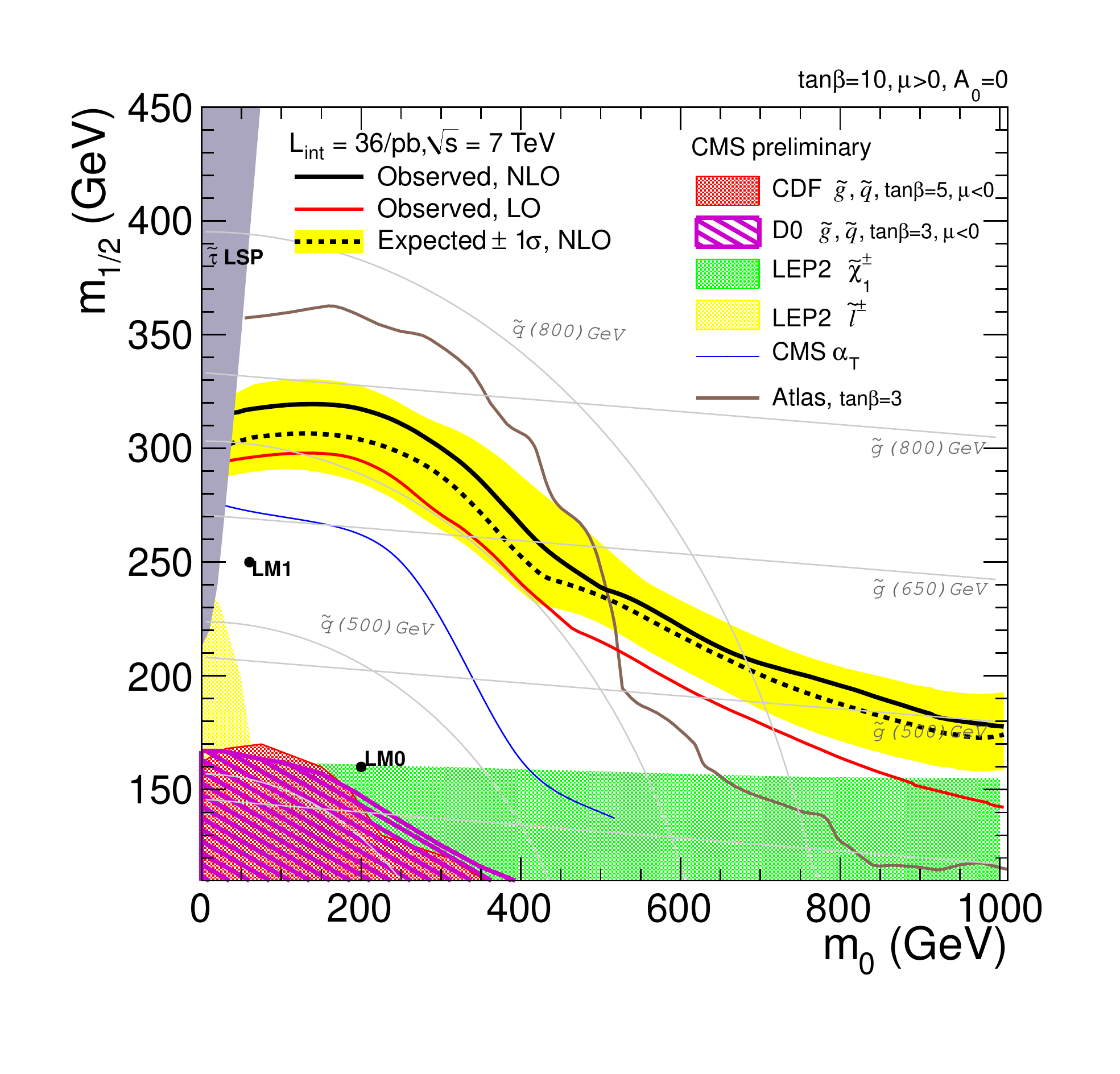}}
\subfigure[]{\raisebox{0.5cm}{\includegraphics[angle=0,width=0.49\textwidth]{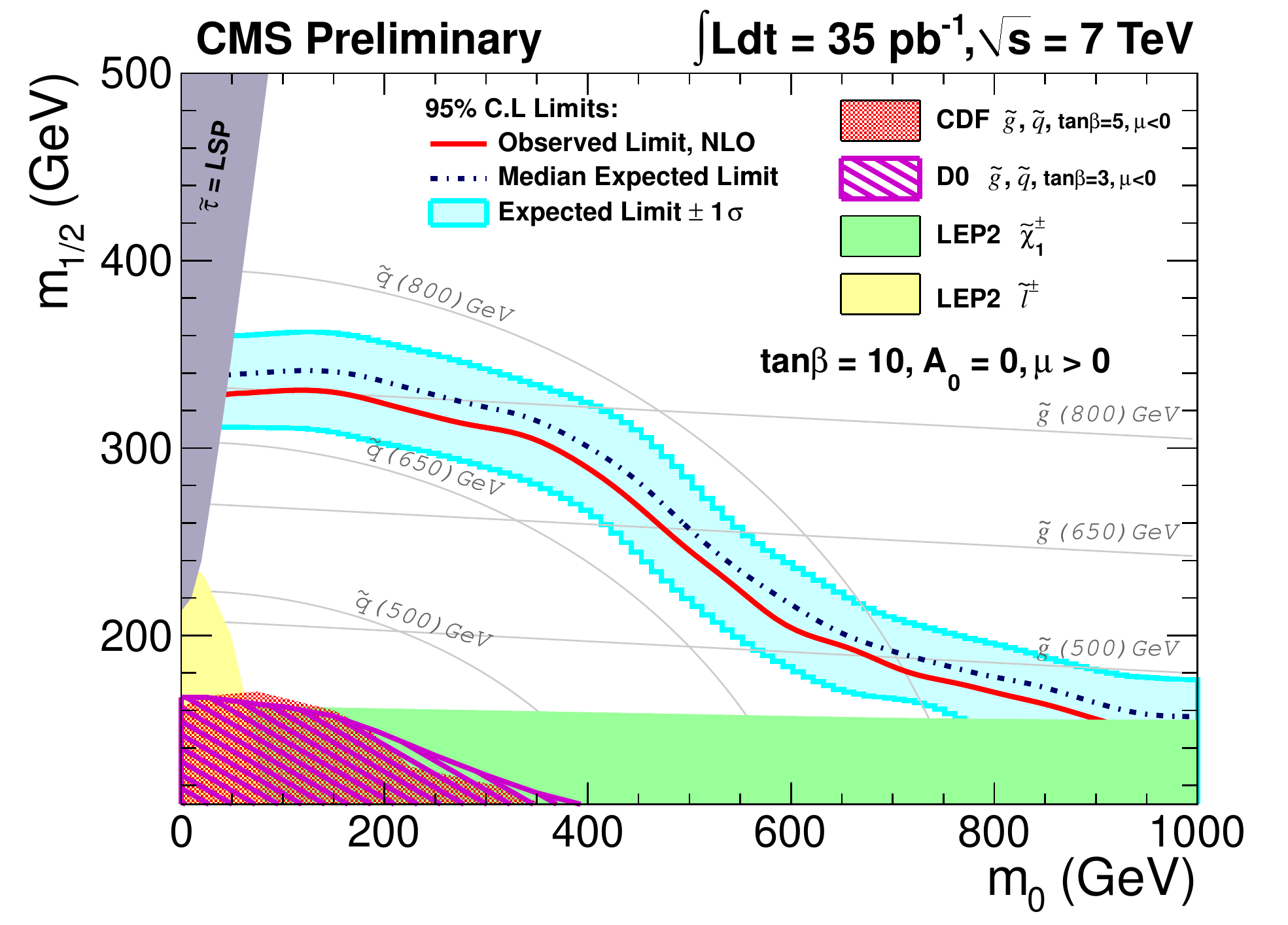}}}\\
\hspace{-1.1cm}
\subfigure[]{\includegraphics[width=0.45\textwidth]{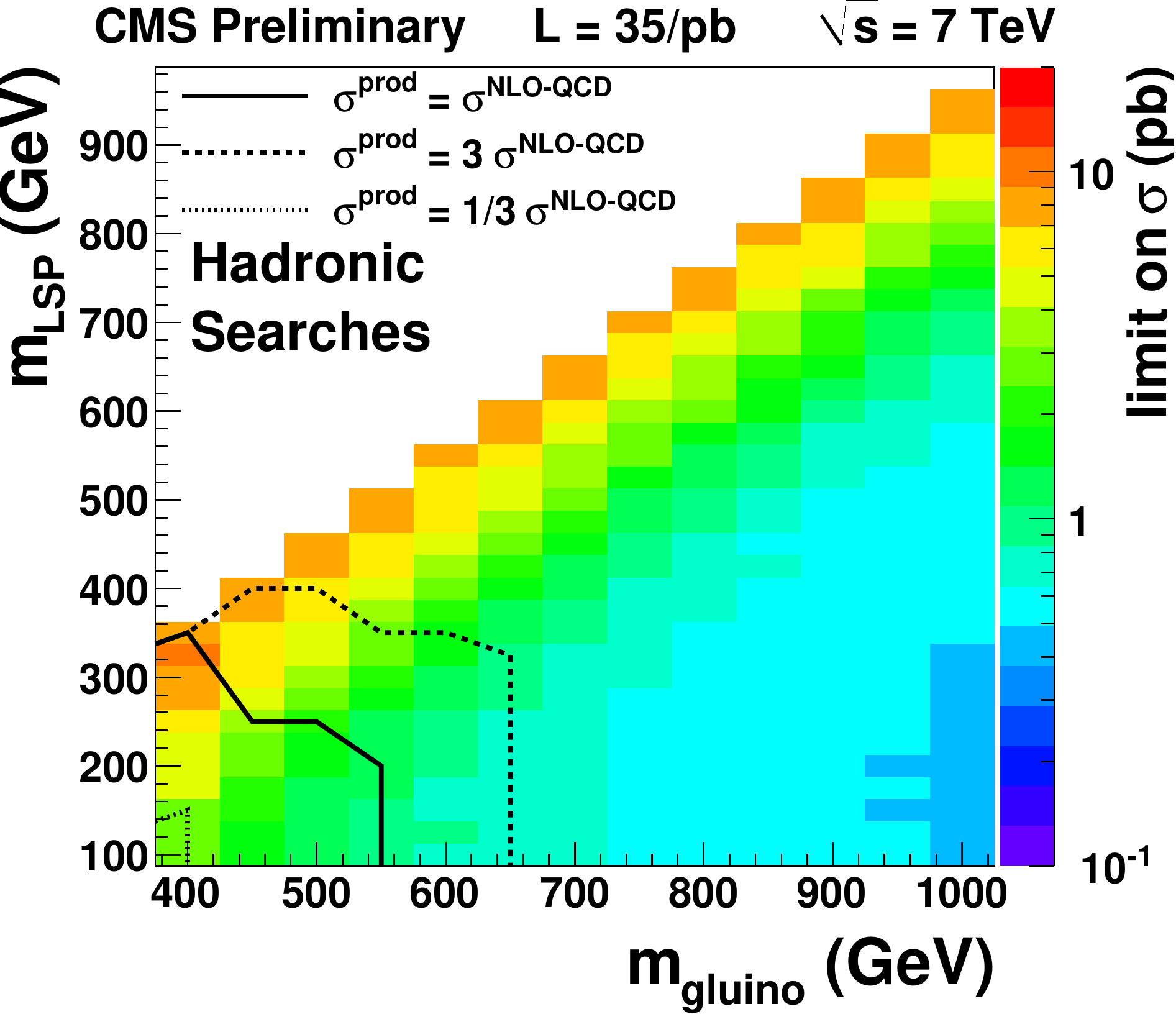}}\hspace{0.5cm}
\subfigure[]{\includegraphics[width=0.45\textwidth]{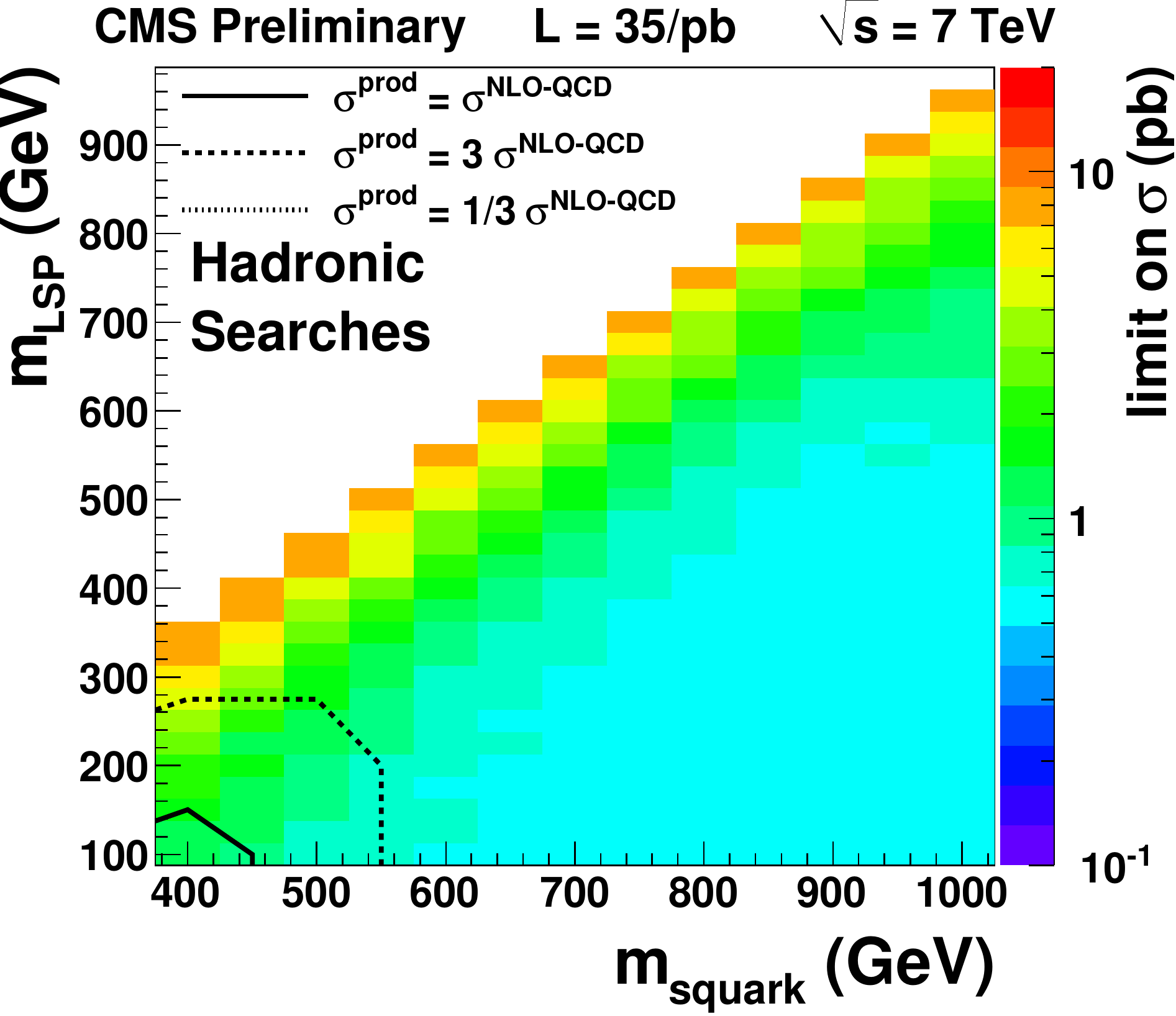}}
\caption{Expected and observed $95\%$ C.L. limits in the cMSSM $(m_0,m_{1/2})$ parameter plane for
(a) the jets+\MHT analysis and (b) the razor analysis. Limits on the di-gluino (c) and di-squark (d) cross-sections in simplified models, obtained by combining the three fully hadronic analyses, namely $\alpha_T$, jets+\MHT and razor. 
%The yellow $1\sigma$-uncertainty band corresponds
%to the expected limit.  The shown contours are the combination of the \HT and the \MHT selection
%such that the contours are the evelope with respect to best sensitivity. 
%The cMSSM parameters are  $\tan\beta=10$, $\mu>0$, and $A_0=0$. {\bf could think about dropping the single lepton figure, which does not bring anything to the global picture.}
}
\label{fig:cMSSM1andSMS}
\end{figure}

After fixing $A_0$,  ${\rm tan} \beta$ and the sign of $\mu$, the model independent upper limit $s^*$ on the number of signal events $s$ from each analysis is projected on the $(m_0,m_{1/2})$ plane by excluding the model if $s(m_0,m_{1/2})>s^*$. The various sources of uncertainty on the signal yield and the signal contamination of the control samples are taken into account. Figures~\ref{fig:cMSSM1andSMS}(a) and (b) present the limits set by the jets+\MHT and razor analyses. 
% 
%\begin{figure}[htb]
%  \begin{center}
%    \includegraphics[width=0.45\textwidth]{Pictures/Susy/T1_Lim_NoThUnc_logZ_combined.pdf}
%    \includegraphics[width=0.45\textwidth]{Pictures/Susy/T2_Lim_NoThUnc_logZ_combined.pdf}
%    \caption{Limits on the di-gluino (a) and di-squark (b) cross-sections in simplified models, obtained by combining the three fully hadronic analyses, namely $\alpha_T$, jets+\MHT and razor. }
%    \label{fig:SMS}
%  \end{center}
%\end{figure}
%
The expected limits are obtained by taking the median of the background test statistics as the result of the experiment, and the $\pm 1  \sigma$ band by taking the median $\pm1\sigma$.

The results of the fully hadronic analyses were also interpreted in the context of two benchmark simplified models~\cite{Alwall-1}: gluino-LSP production (left) and squark-LSP production (right).  
The former refers to pair-produced gluinos, where each gluino directly decays to two light quarks and the LSP resulting in a four jet plus missing transverse energy final state.
The latter refers to pair-produced squarks, where each squark decays to one jet and the LSP resulting in a two jet plus missing transverse energy final state. Figures~\ref{fig:cMSSM1andSMS}(c) and (d) show the upper limit on the cross-section as a function of the physical masses of the particles involved in each model. In each bin, the upper limits obtained in the $\alpha_T$, the jets+\MHT and the razor analyses are considered, and the minimum one is shown. Theoretical uncertainties are not included.

\section{Leptonic searches}

The single lepton analysis~\cite{ref:CMSSingleLepton} selects events featuring jets, \MET, and a single lepton in the final state. 
The presence of the lepton strongly reduces the contribution of the QCD multi-jet and \Znunujets backgrounds, and provides several handles to build a data-driven prediction of the remaining background contribution from QCD, \ttbar, and \Wjets. Events containing an additional lepton are vetoed, and handled by the di-lepton and multi-lepton analyses. 
The event sample was collected using triggers based on the presence of a single electron or a single muon. The requirement of an \HT trigger was added when the peak luminosity increased beyond $2 \times 10^{32}\,\cm^{-2}s^{-1}$. The trigger selection is fully efficient with respect to the baseline selection applied offline, which consists of requiring (i) four jets with  $\pt > 30$\,\GeVc and $|\eta|<2.4$ with $\HT>500\,\GeV$; (ii) an isolated lepton, which can be either a muon with $\pt > 20$\,\GeVc and $|\eta|<2.1$, or an electron with $\pt > 20$\,\GeVc and $|\eta|<2.4$. The search region is defined by an additional cut on the missing transverse energy, $\MET>250\,\GeV$.
The contribution of the main background processes to the search region, \ttbar and \Wjets, is estimated using the lepton spectrum method. The foundation of this method is that, when the lepton and neutrino are produced together in a $W$ decay (either in $t\bar t$ or in $W$+jets events), the lepton \pt spectrum is directly related to the neutrino \pt spectrum.  The lepton spectrum is used to predict the \MET distribution, after suitable corrections related to the effect of the $W$ polarisation on the lepton and neutrino \pt spectra, and to the lepton acceptance and reconstruction efficiency.
Combining the electron and muon channels, 2 events are observed in the search region,  while $3.6 \pm 2.9$ are expected. A 95\% model independent upper limit of 4.1 signal events is calculated. In the cMSSM, for ${\rm tan} \beta=10$, $A_0=0$, and $\mu>0$, gluino and squark masses larger than about 550\,\GeVcc are excluded. 

The same-sign di-lepton analysis requires, in addition to jets and \MET, exactly two isolated leptons of the same sign which can be electrons, muons or taus decaying hadronically. The event sample was collected using di-lepton and single-lepton triggers, but also \HT triggers, which provide sensitivity to events with low \pt electrons and muons. The search selection and the data-driven background estimation techniques employed where chosen according to the trigger in use (lepton or hadron), and the channel ( $l_i l_j$ where $l_{i,j} = e, \mu,\tau$). In all search regions, the predicted number of background events is compatible with zero, and no excess is observed. The analysis and the results are described in details in Ref~\cite{ref:CMSSameSign}, which also provides lepton efficiency maps that can be used to test a variety of models.  

The multi-lepton analysis~\cite{ref:CMSMultiLeptons} selects events with three isolated leptons or more, acquired using single-lepton and di-lepton triggers. The events are sorted in 54 independent samples according to the relative charge of the leptons and their flavour, which can be $e, \mu$, and $\tau$. The three-lepton requirement strongly reduces the standard-model background, and the largest remaining background process is \Zjets, including Drell-Yan. The remaining background is further suppressed by requiring $\HT>30\,\GeV$, $\MET>50\,\GeV$ or a $Z$ veto, depending on the considered final state. No excess is found with respect to the predicted background in search region, and limits are set in a variety of models. In particular, in the so-called co-NLSPs scenario (see Ref~\cite{Ruderman:2010kj} and references therein), squark and gluino masses lower than 830\,\GeVcc and 1040\,\GeVcc are excluded. 

\section{Conclusion}

Complementary searches for Supersymmetry and other new physics leading to similar final states were conducted at CMS using the 35\,\pbinv of data collected in 2010, in a wide variety of final states. No excess has been observed so far with respect to the expectations from the standard model, and stringent limits were set in various SUSY models.
Data-driven background estimation techniques have been used wherever possible, paving the way towards the analysis of high-luminosity 2011 data. 

\section{Acknowledgements}

I would like to thank the members of the CMS collaboration for the excellent performance of the detector and of all the steps culminating in these results, as well as the members of the CERN accelerator departments for the smooth operation of the LHC machine. 

%\section*{References}
%\begin{thebibliography}{99}
%\bibitem{ja}C Jarlskog in {\em CP Violation}, ed. C Jarlskog
%(World Scientific, Singapore, 1988).

%\bibitem{ma}L. Maiani, \Journal{\PLB}{62}{183}{1976}.

%\bibitem{bu}J.D. Bjorken and I. Dunietz, \Journal{\PRD}{36}{2109}{1987}.

%\bibitem{bd}C.D. Buchanan {\it et al}, \Journal{\PRD}{45}{4088}{1992}.

%\end{thebibliography}

\bibliography{colinBernet}

\end{document}